\documentstyle[prb,aps,twocolumn,floats]{revtex}
\begin{document}

\title {\bf Universality in an integer Quantum Hall transition}

\author{ P.T. Coleridge }

\address
{Institute for Microstructural Sciences, National Research
Council, Ottawa, Ontario, K1A OR6, Canada\\ }

\date{5 February 1999}

\maketitle
%\begin{abstract}
%\small
{\bf
 An integer Quantum Hall effect transition is studied in a 
modulation doped p-SiGe sample. In contrast to most examples
of such transitions the longitudinal and Hall conductivities
at the critical point are close to 0.5 and 1.5 (e$^2$/h),
the theoretically expected values. This allows the 
extraction of a scattering parameter, describing both 
conductivity components, which depends exponentially on filling
factor. The strong similarity of this functional form to those
observed for transitions into the Hall insulating state and for the
B=0 metal-insulator transition implies a universal quantum critical 
behaviour for the transitions. The observation of this behaviour in the 
integer Quantum Hall effect, for this particular sample, is attributed to 
the short-ranged character of the potential associated with the dominant
 scatterers. 

}
%\end{abstract}

%\newpage

     Despite the precision with which Integer Quantum Hall (IQH)
effect plateaus are quantised the behaviour between plateaus is not
yet well understood. Two recent review articles
\cite{DasSarma,Sondhi} have pointed out that for these continuous
quantum phase transitions the predictions of theory are only
partially confirmed by experiment. Some of these predictions are:
 a peak value of the conductivity between IQH states
$\sigma_{xx}^c$ with a universal value of e$^2$/2h \cite{Huo,DHL},
temperature scaling behaviour showing a $T^\kappa$ 
dependence with $\kappa$ probably close to 3/7 \cite{Huckestein} and
the two components of the conductivity connected by a semi-circular
relation \cite{DHL,Ruzin}. Experimentally, in samples showing the
best scaling behaviour values of  $\sigma_{xx}^c$ are less than 
e$^2$/2h \cite{Wei} and where values of
 $\sigma_{xx}^c$ are close to 0.5  scaling does not usually appear
to work \cite{DasSarma}. For transitions from the high field Hall 
insulator (HI) phase, into the $\nu$ =1 plateau, or the FQHE
 $\nu$ = 1/3 plateau the situation is somewhat better. There are
now several papers\cite{Shahar95,Wong} which find transitions that 
appear to scale well,  with the expected critical values for $\sigma_{xx}$ 
and $\sigma_{xy}$. 

     Results are presented here for a $\nu$ = 2 to 1 IQH transition
which conform, in many ways, to the theoretical expectations. They are
obtained in a strained, modulation doped p-SiGe sample with
properties that differ, in a number of ways, from the usual GaAs
based heterojunctions. This points to some of the factors that usually
suppress the theoretically predicted behaviour. An
interpretation of these results is offered which supports the
universal nature of the transition.

      The sample \cite{sample} has a density of 
3.4 $\times$ 10$^{15}$ m$^{-2}$, a transport mobility ($\mu_{tr}$) of
 1.3 m$^2$/Vs and a quantum mobility ($\mu_q$) of 1.5 m$^2$/Vs .
Although these mobilities appear rather low, the quantum lifetime
($\tau_q$ = 2.4ps) is in fact equal to that of a GaAs
based 2D-electron gas (2DEG) with a mobility over 100 m$^2$/Vs. 
This is because in GaAs the effective mass is
significantly smaller (.067m$_e$ compared with
0.27m$_e$ here) and the ratio $\mu_{tr} / \mu_q$  is typically 20. 
The value of about one seen here implies the dominant scattering mechanism 
in p-SiGe samples involves a short-ranged potential.

      The holes reside in the J = 3/2, $|M_J|$ = 3/2, heavy hole band,
split-off from the light hole band by strain and confinement \cite{sample}.
In low magnetic fields the spin-splitting (or more correctly
 parity splitting)is large, approximately 70 \% of the cyclotron
spacing.  At higher fields there is a strong exchange enhancement
of the splitting so at filling factor $\nu$ = 2 the (1$\uparrow$) and
 (0$\downarrow$) levels have crossed (see figure 1) and the system becomes
ferromagnetically polarised \cite{ssc}.  Activation measurements show the
spin-splitting at $\nu$ = 2 is almost twice the cyclotron spacing so the 
(1$\uparrow$) Landau level, involved in the $\nu$ = 2 to 1 Quantum
 Hall transition, is very well removed from adjacent Landau
levels \cite{footnote1}.

	Figure 1 shows $\rho_{xx}$ data at several temperatures 
for the $\nu$ = 2 to 1 IQH transition  obtained in a 200$\mu$m wide
Hall bar with a measuring current of 1nA. There is a well defined
fixed point on the low field side of the peak and inverting the
resistivity data (see figure 2a) shows this point corresponds to
$\sigma_{xx}$ = 0.46 and $\sigma_{xy}$ = 1.52 (in units of e$^2$/h),
close to the expected values of 0.5 and 1.5. Figure 3
shows a scaling plot of the data from figure 1 collapsed onto a
single curve by plotting it against ($\nu_c - \nu$)/T$^{\kappa}$.
This involves no adjustable parameters, the critical value $\nu_c$ 
( = 1.612) is determined by the field at the fixed point and the 
exponent $\kappa$ (= 3/7)has been taken as the generally accepted 
theoretical value \cite{Huckestein}. Scaling behaviour is obtained 
for temperatures between 70 and 180mK, a comparable quality of fit 
is obtained for  $\kappa$ varying by about 10\%. At higher 
temperatures deviations, illustrated by the 400mK data, occur
 simultaneously in the peak height, in the peak width, and in the 
movement of the ''fixed point`` where the curve crosses the collapsed data. 
Below 70mK the width changes little, 
consistent with an effective temperature limited to about 60mK. 
It is not clear whether this is a genuine effect or just reflects an 
experimental inability to cool the carriers significantly below this 
temperature.

      Figure 2b shows $\sigma_{xx}$ plotted against $\sigma_{xy}$ with 
the predicted semicircular behaviour\cite{DHL,Ruzin} shown by a 
dashed line. There is a general deviation of approximately 
10 \% between the two curves. The model used in Ref. \onlinecite{DHL}
suggests the conductivity components (for the $\nu$ = 2 to 1
transition) should be given by

\begin{equation}
            \sigma _{xx} = s / (1 + s^{2})   \; \; \; \; \; \; \; \;
                 \sigma _{xy} = 2 - s^{2} / (1 + s^{2})                       
\label{eq1}
\end{equation}

where s, a parameter that varies from 0 to $\infty$ through the
transition, is equal to 1 at the critical point. It can be
identified with $\sigma _{xx} ^{(b)}$ in the Chern-Simons mapping
or equivalently with the ratio R/T where R and T are respectively
quantum percolation reflection and transmission coefficients. 
Figure 4 shows values of s extracted from both 
$\sigma_{xx}$ and $\sigma_{xy}$ using these expressions. For the
data derived from $\sigma_{xy}$ (solid line) ln(s) varies
approximately linearly with the filling factor
 $\delta \nu = \nu_c - \nu$. Well away from $\delta \nu$ = 0 the
noise, and maybe some of the deviations from linear behaviour,
should be attributed to the fitting procedure whereby s$^2$ is
fitted to the small deviations of $\sigma_{xy}$ from quantised
plateau values. 

      For the values derived from $\sigma_{xx}$ (dashed lines) there
is a discontinuity around $\delta \nu$  = 0 because of the small
deviation of the peak value of $\sigma_{xx}$ from the exact value of 0.5.
 Away from $\delta \nu$ = 0, however, there is a general agreement 
between the two separately determined values of s. Renormalising the 
$\sigma_{xx}$ data to the 0.5
at $\delta \nu$ = 0 produces excellent agreement between the 
two values of s near $\delta \nu$ = 0 without any significant change
for larger vbalues of $\delta \nu$.  A feature of the $\sigma_{xx}$ 
results is the very high degree of inversion symmetry. This is
demonstrated in fig. 4, by simultaneously plotting s against
$\delta \nu$ and 1/s against -$\delta \nu$ on the same graph where it
can be seen that $s(\delta \nu ) = 1/s(-\delta \nu)$ over a wide
range of s.

      The other important feature of the results in fig.4 is the
essentially linear dependence of ln(s) on $\delta \nu$ around
 $\delta \nu$ = 0. Combined with the scaling observed in figure 3 this 
implies, for small values of $\delta \nu$

\begin{equation}
             s     =  \exp [ (\nu_c -\nu) (T_0 /T)^{\kappa}],  
\label{eq2}
\end{equation}

with $T_0$ a constant, approximately equal to 280K. This behaviour
is not consistent with the procedure used in Ref. \ \onlinecite{DHL} 
to relate s to
$\delta \nu$ but is reminiscent of the model proposed by
Dobrosavljevic et al\cite{DAMC} to describe the B = 0 metal-
insulator transition in Si-MOSFETs\cite{krav} where it explains
naturally the inversion symmetry observed between $\rho_{xx}$ and
1/$\rho_{xx}$ on the two sides of the transition\cite{Simonian}.
  A similar expression, but with a different temperature dependence,
 has also been proposed by Shahar et al\cite{Shahar98} to explain 
the IQH-Hall Insulator transition in GaAs based samples.
In this case the symmetry around $\delta \nu$ = 0 is observed 
 in the I-V curves\cite{Science} as well as in $\rho_{xx}$. 

      The connection between the IQH 1-2 transition and the IQH-HI
transition is seen very clearly if the formalism outlined in eqns
1 and 2 is applied to Hall insulator transition, but
with $\sigma_{xy} = 1 - s^2 / (1+s^2)$ . The resistivities
obtained by inverting $\sigma_{xx}$ and $\sigma_{xx}$ are then just
$\rho_{xx}$ = s and $\rho_{xy}$ = 1. The exponential dependence of
$\rho_{xx}$ on $\delta \nu$ found in Ref. ~\onlinecite{Shahar98} is 
therefore the same as  that in Eqn.2 indicating the very direct
 equivalence between the two transitions\cite{footnote2}. A very similar 
equivalence has previously been noted by Shahar et al.\cite{Shahar97}
although without an explicit evaluation of s. Incorporating an E field 
dependence into Eqn. 2 can also empirically account for the symmetry 
observed in the I-V curve. In particular for the absence of any I-V 
dependence in the $\rho_{xy}$ data for this transition\cite{Science}. 

      The specific functional dependence of s, indicative of a
continuous quantum phase transition, therefore explains both the 
 transition between IQH phases and the IQH-HI transition. Further, if
s is identified with the resistivity at B = 0 (and $\delta \nu$ with
 the density changes) it is also consistent with the behaviour of the
 B=0 metal-insulator transition \cite{DAMC}.
 These transitions all therefore  appear to belong to the same
universality class. There remains, however, the problem of the
residual discrepancy of about 10\% between the observed and
expected values of $\sigma_{xx}$ and the related question of why, 
in most samples, the discrepancy is so much larger. These large 
discepancies are presumeably the reason the intrinsic properties of the
 phase transition are not always revealed in the transport data.

      The p-SiGe system differs from the more common GaAs and GaInAs 
based systems in a number of ways that may be relevent to this issue. 
 They include: approximately equal values of $\tau_q$ and
$\tau_{tr}$ indicating a short-ranged scattering potential;
spin resolved Landau levels that are well separated;
 a spin polarised system; and an IQH transition occurring in the
 $N_L$ = 1, not $N_L$ = 0 Landau level.

	The data reported in Ref.\onlinecite{Shahar97} show a $\nu$ = 2 to 1
IQHE transition very similar to that reported here. The peak value of 
$\sigma_{xx}$ is about 0.45 and an approximately semicircular relationship
between $\sigma_{xx}$ and $\sigma_{xy}$ is also observed. The sample there 
was a low mobility GaAs based 2DEG: a value of $\mu_q$ is not available but
the low  mobility, for relatively high density, implies the dominant 
scatterers are almost certainly impurities in the channel. Unlike remote
donor impurities these have a short-ranged scattering potential. The
common feature between these two samples is the short-ranged potential. 
This is therefore, almost certainly, the reason for the similar,
almost canonical, behaviour of the quantum Hall transition in both cases.

      In high mobility GaAs based 2DEGs, where the scattering potential 
is long ranged, the ratio $\tau_{tr}/\tau_{q}$ is large because the 
momentum weighting term (1-cos$\theta$) is small. In these circumstances 
the quantum diffusion model gives quantitatively correct 
answers\cite{pkval}. It predicts peak values
 of $\sigma_{xx}$ that are reduced by a factor of approximately
 $\tau_{q}/\tau_{tr}$ from those given for simple $\delta$-function
 scattering\cite{Ando}. This ratio represents the difference between 
the lifetime $\tau_{q}$ , that characterises the Landau level width 
(and hence the peak value of the density of states) and the transport 
time $\tau_{tr}$ which includes the condition that multiple small 
angle scattering events are needed to randomise the velocity of the 
carriers.

      In the same spirit, in IQH transitions when localisation
effects are important, although the scaling behaviour may be dominated
 by the band of itinerant states near the Landau level centre the observed
conductivity will also depend on the transport process and the range of the 
scattering. For example, in quantum percolation models the conductivity is
determined by quantum amplitudes that depend on the degree of tunnelling at
 saddle points \cite{DHL,CC}. Explicitly it is obtained from these 
amplitudes by using a one dimensional, edge state, model where there is 
exact cancellation between the density of states and the velocity ($v_F$). 
In practice this model will not always be valid and the transport 
will remain, to some extent, 2-dimensional. The component of velocity that 
determines the conductivity will then take a range of values, not just 
$\pm v_F$. For $\tau_q \approx\tau_{tr}$ , the exact one-dimensional 
cancellation between the density of states and the velocity will be 
approximately reproduced and the conductivity should then reflect quite 
accurately the behaviour of the quantum amplitudes and the associated 
universality. Generally, however, there is a ``memory'' associated with 
the multiple scattering events that are needed to randomise the carriers 
when scattered by a long-range potential and this reduces $\sigma_{xx}$ 
below the "universal" value. This reduction, like that seen in the quantum 
diffusion model, depends on the range of the potential and will be
sample dependent.

	The calculations in Ref. \onlinecite{Huo} also give universal 
values of $\sigma_{xx}^c$ but using a different approach. They were  
restricted to potentials with ranges less than the magnetic length.
For example in a field of 9 tesla (such as that for the IQH 
transition discussed here) the magnetic length is about 8nm whereas the 
range of the potential in clean,modulation doped, 2DEGs is roughly 
twice the set-back distance, typically more than 80nm. 

      In summary, an integer quantum Hall effect transition measured
in a relatively high mobility 2-D system is found to conform rather
closely to theoretical predictions. A scattering parameter is
extracted which is exponentially dependent on the filling
factor and which has a very high degree of symmetry about the
critical point. Application of the same formalism to Quantum Hall -
 Hall insulator transitions explains, naturally , the dualities
observed there and also emphasises similarities to the B=0 metal
insulator transition. It is suggested that failure to observe this
canonical behaviour in most Quantum Hall transitions is associated
with the conductivities only revealing the underlying
properties of the phase transition if the scattering is dominated
by short range potentials.

	The author would like to thank Dr. R. Williams for material growth,
 Dr. Y. Feng for sample preparation and Dr. A.S. Sachrajda and P. Zawadzki 
for use of the dilution refrigerator and for assistance with the 
measurements. Also Professors S. Das Sarma, D.-H. Lee
and S.L. Sondhi for helpful discussions.

\newpage

\begin{figure} [p]
\vspace*{7.3cm}
\includegraphics{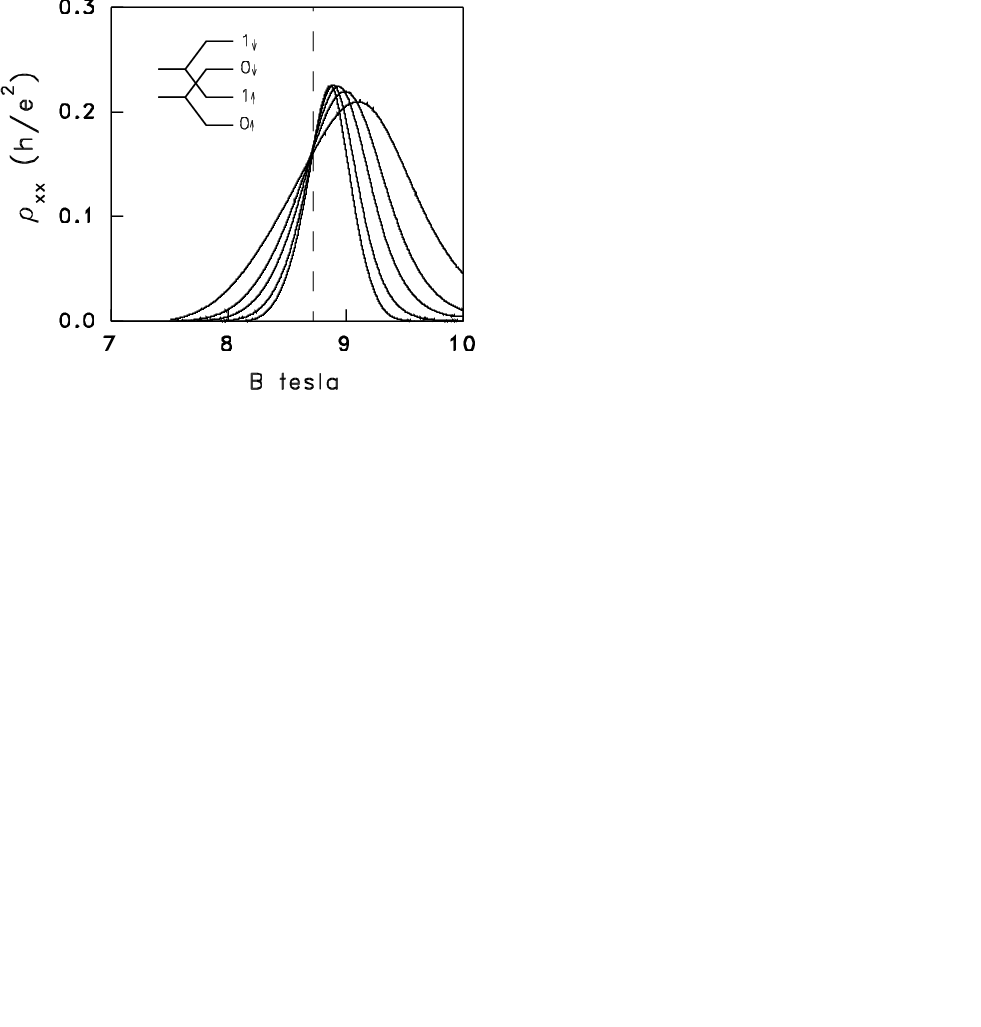}
%\special{eps:fig1_sm.eps x=7cm y=6cm}
\caption
{Resistivity data ($\rho_{xx}$) for the $\nu$ = 2 to 1 IQH effect transition
at temperatures of 70,120,180,250 and 400 mK. The inset shows, schematically
the ferromagnetic alignment of the spins at $\nu$ = 2. }
\label{fig1}
\end{figure}

\begin{figure} [p]
\vspace{9.3cm}
\includegraphics{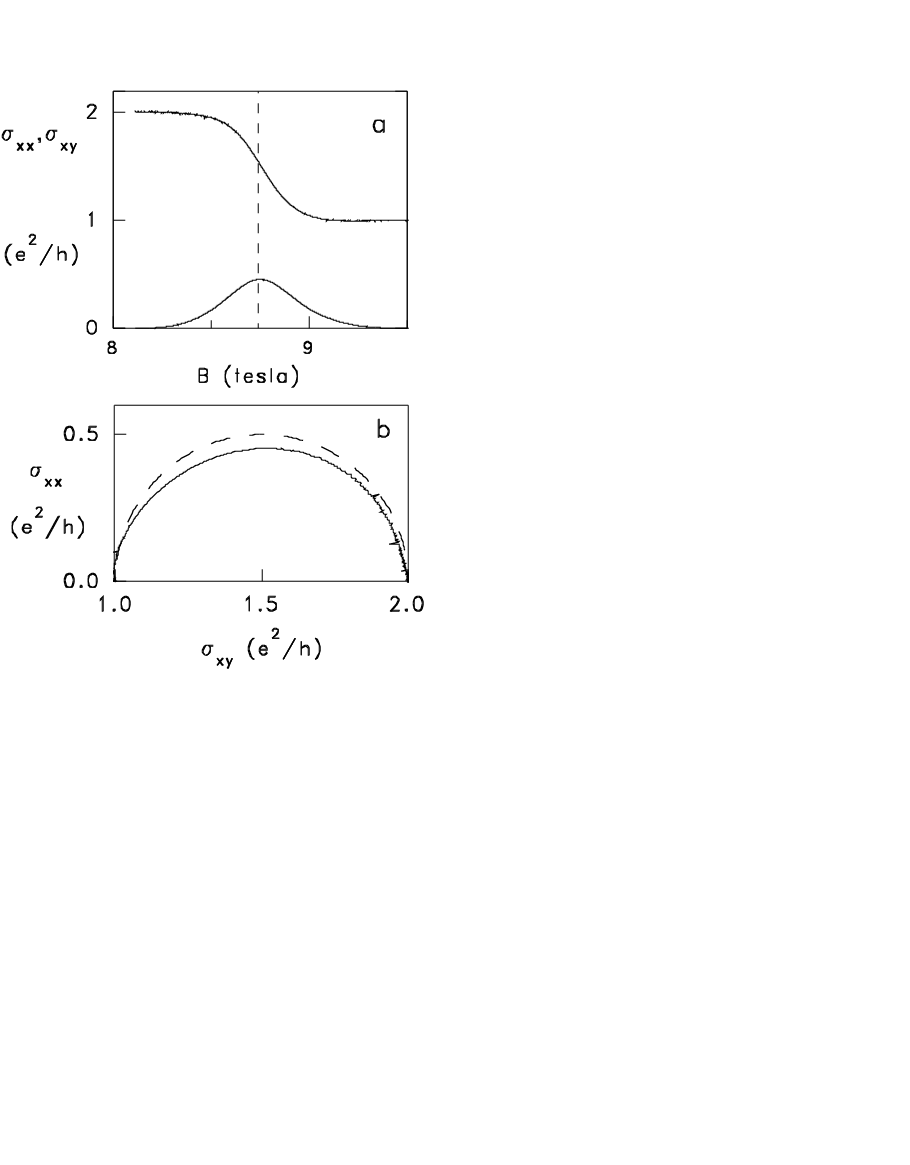}
\caption
{(a) Conductivity data obtained by inverting resistivities measured at
 approximately 65mK. The dashed line shows the critical field 
indicated in figure 1. 
(b) $\sigma_{xx}$ plotted against $\sigma_{xy}$ with the semicircular 
	relationship, $\sigma_{xx}^2 + (\sigma_{xy} - 3/2 )^2 = 1/4 $ 
shown, dashed,for comparison. }
\label{fig2}
\end{figure}

\begin{figure} [p] 
\vspace{8.0cm}
\includegraphics{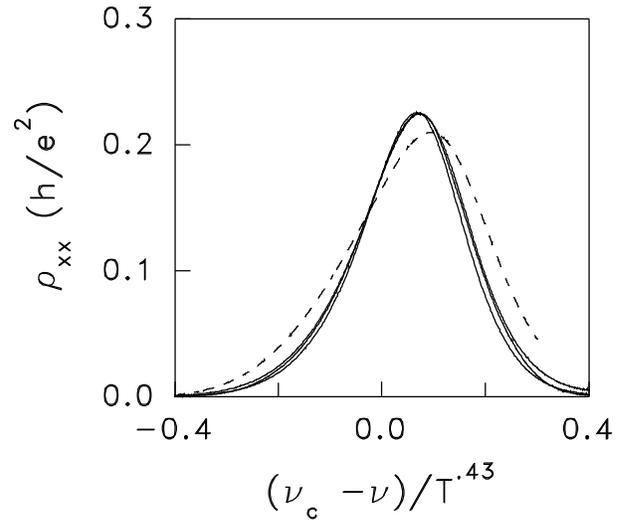}
\caption
{Scaling curves showing the resistivity data from fig. 1, for 
T = 70,120,180mK,
plotted against $(\nu_c - \nu)/T^{\kappa}$ with T in kelvins,
$\kappa$ = 3/7 and $\nu_c$, determined from fig.1, equal to 1.612.
	The dashed line is 400mK data.
 }
\label{fig3}
\end{figure}

\begin{figure} [p]
\vspace{8.0cm}
\includegraphics{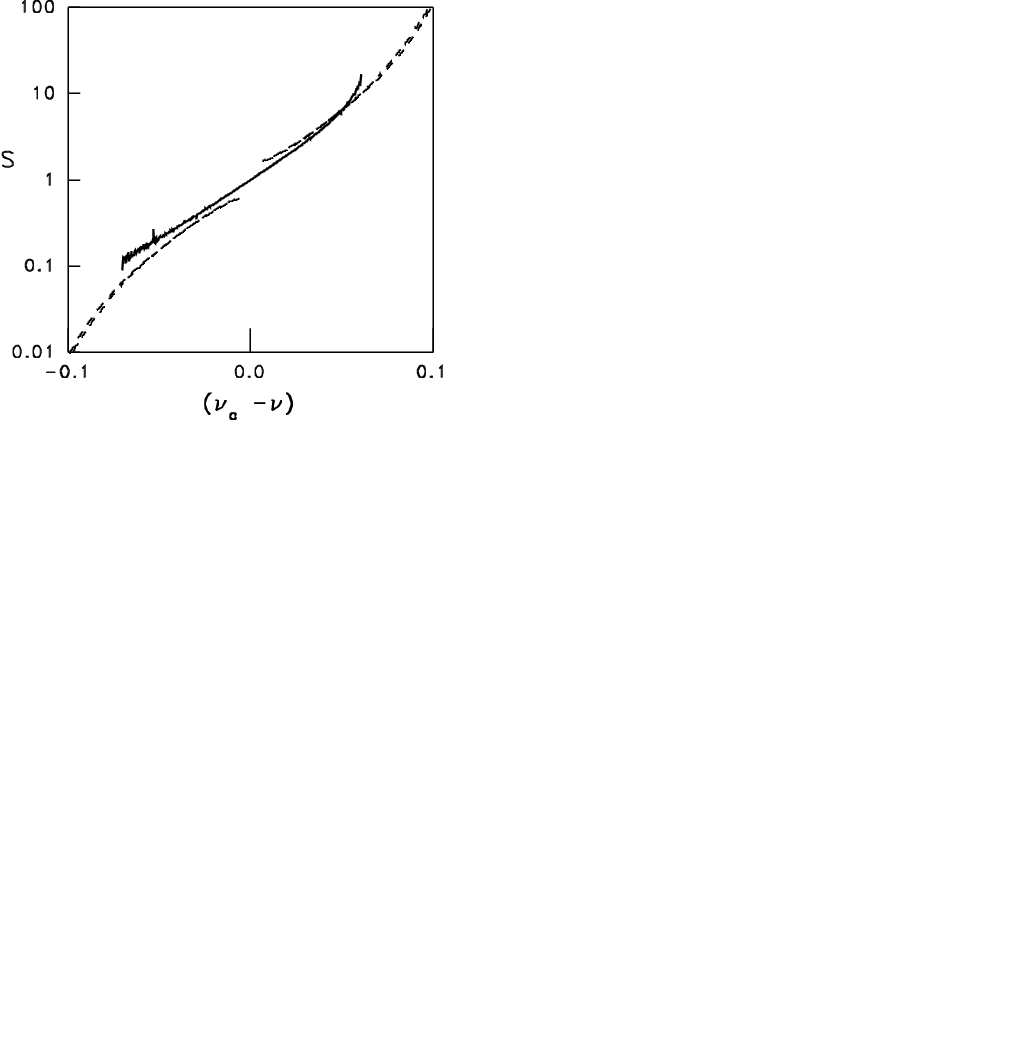}
\caption
{Scattering parameter s, as defined in the text, determined from 
the components of the conductivity plotted on a logarithmic scale.
 Solid line, derived from$\sigma_{xy}$; dashed line, derived from
$\sigma_{xx}$, plotted both as s against $\delta \nu$ and as 1/s
 against -$\delta \nu$.
}
\label{fig4}
\end{figure}


\begin{references}
\bibitem{footnote1}
      As the field is increased, however, the (1$\uparrow$) and 
(0$\downarrow$)levels approach because the exchange enhancement, relative 
tothe cyclotron spacing, is reduced. In similar, but less dense samples, 
the reduction is sufficient to cause the two levels
to actually overlap near  $\nu$ = 1.5 . The small
deviations near $\sigma_{xy}$ = 1, observed in fig2b, can probably be
associated with this phenomenon.
\bibitem{footnote2}
	It should be noted that over the temperature range for which a single
crossing point is observed,  the temperature dependence seen in Fig. 1 
which is explained by scaling with $\kappa$ = 3/7, is also 
consistent (within experimental error) with a linear T dependence
like that observed in Ref. \onlinecite{Shahar98}.
\end{references}
\end{document}